\documentclass[paper]{agujournal2019}
\usepackage{url}
\usepackage{lineno}
\usepackage{color}
\usepackage{amssymb,amsmath}
\usepackage{graphicx}
\usepackage{makecell}
\draftfalse

\newcommand{\ssr}{   {Space Science Review}}

\newcommand{\jgr}{   {J. Geophys. Res.}}
\newcommand{\grl}{   {Geophys. Res. Lett.}}

\newcommand{\apjl}{   {Astrophys. J. Lett.}}

\newcommand{\nat}{   {Nature}}
\newcommand{\aap}{   {Astronomy \& Astrophysics}}

\journalname{JGR: Space Physics}

\begin{document}


\title{Variation of Whistler-Mode Wave Characteristics Along Magnetic Field Lines: Comparison of Near-Equatorial THEMIS and Middle-Latitude ERG Observations}

\authors{Sophie Kadan\affil{1}, Xiao-Jia Zhang \affil{2,3}, Anton Artemyev\affil{3}, 
Yoshizumi Miyoshi\affil{4}, Ayako Matsuoka\affil{5}, Yoshiya Kasahara\affil{6}, Shoya Matsuda\affil{6}, Tomoaki Hori\affil{4}, Mariko Teramoto\affil{7},
Kazuhiro Yamamoto\affil{4}, Iku Shinohara\affil{8}
} 
\affiliation{1}{University of Pennsylvania, Philadelphia, PA, USA}
\affiliation{2}{Department of Physics, University of Texas at Dallas, Richardson, TX, USA}
\affiliation{3}{Department of Earth, Planetary, and Space Sciences, University of California, Los Angeles, USA}
\affiliation{4}{Institute for Space Earth Environmental Research, Nagoya University, Nagoya, Japan}
\affiliation{5}{Graduate School of Science, Kyoto University, Kyoto, Japan}
\affiliation{6}{Graduate School of Natural Science and Technology, Kanazawa University, Kakuma, Kanazawa, Japan}
\affiliation{7}{Kyushu Institute of Technology, Faculty of Engineering, Department of Space Systems Engineering, Kitakyushu, Fukuoka, Japan}
\affiliation{8}{Institute of Space and Astronautical Science, Japan Aerospace Exploration Agency, Sagamihara, 252-5210, Kanagawa, Japan}

\correspondingauthor{Sophie Kadan}{sokadan@sas.upenn.edu}

\begin{keypoints}
\item We present near-equatorial THEMIS and off-equatorial ERG measurements of whistler-mode waves from the same MLT sector
\item The comparison of wave spectrum characteristics confirms that THEMIS and ERG observe the same wave activity
\item We validate the wave empirical models by comparing the wave intensity measurements from THEMIS and ERG
\end{keypoints}

\begin{abstract}
The latitudinal distribution of whistler-mode wave intensity plays a crucial role in determining the efficiency and energy of electrons scattered by these waves in the outer radiation belt. Traditionally, this wave property has mostly been derived from statistical measurements of off-equatorial spacecraft, which collect intensity data at various latitudes under different geomagnetic conditions and at different times. In this study we examine a set of events captured by both the near-equatorial THEMIS spacecraft and the off-equatorial ERG spacecraft. Specifically, we compare the whistler-mode wave intensity from THEMIS and ERG measurements at the same MLT and time sectors. Similar wave spectrum characteristics confirm that THEMIS and ERG indeed observed the same wave activity. However, upon closer examination of the wave intensity variations, we identify two distinct categories of events: those that follow the statistically predicted variations in wave intensity along magnetic latitudes, and those that exhibit rapid wave intensity decay away from the equatorial plane. We analyze main characteristics of events from both categories and discuss possible implications of our analysis for radiation belt models.
\end{abstract}

\section{Introduction}
Electromagnetic whistler-mode waves represent a key natural wave emission that controls the dynamics of energetic electron fluxes in the outer radiation belt \cite<see>[and references therein]{Thorne21:AGU}. Electron resonant acceleration up to relativistic energies is mostly determined by characteristics of the near-equatorial waves that resonate with large pitch-angle electron populations \cite<e.g.,>{Thorne13:nature,Li14:storm,Mourenas14:fluxes,Allison&Shprits20}. Electron resonant scattering into the loss cone and their subsequent precipitation into the atmosphere is determined by off-equatorial waves that may scatter relativistic electrons \cite<see discussions in>{Miyoshi15:aurora, Miyoshi20,Tsai22,Artemyev21:jgr:ducts,Chen22:microbursts}, whereas near-equatorial waves are usually responsible for precipitation of $<100$keV electrons \cite{Li13:POES,Ni14:POES}. Therefore, the latitudinal distribution of whistler-mode wave intensity largely controls the energy range of electron losses \cite<see discussion in>{Agapitov18:jgr,Wang&Shprits19:latitudes,Miyoshi20,Tsai24}. The statistical parameterization of such latitudinal distributions is thus an important yet unresolved question in studies of the outer radiation belt.

In the last several decades, only two spacecraft missions, Polar \cite{Gurnett95} and Cluster \cite{Escoubet01}, provided regular measurements of off-equatorial (middle latitude) whistler-mode waves \cite<e.g.,>{Haque10,Tsurutani11,Agapitov13:jgr}. Likewise, conjunctions of these measurements with near-equatorial whistler-mode wave measurements are rather limited. An instance involving near-equatorial Van Allen Probes \cite{Mauk13} and middle-latitude Arase \cite<ERG;>{Miyoshi18:ERG} measurements has been illustrated in \citeA{Colpitts20} and \citeA{Santolik21:rbsp&erg} (see also, \citeA{Miyoshi22:ssr}): this study identifies a robust correlation in the wave characteristics detected by two well-separated spacecraft, which motivates a further statistical exploration of near-equatorial and middle-latitude wave measurements. Furthermore, the highly inclined ERG orbits ($\sim 31^\circ$) almost ideally fulfills the primary requirements for such investigations \cite{Miyoshi18:ERG}. Throughout several months in 2019, 2021, and 2022, the ERG mission conducted measurements of mid-latitude whistler-mode waves in the same MLT sector as the near-equatorial THEMIS \cite{Angelopoulos08:ssr} mission. This offers a unique opportunity for a statistical comparison of wave intensities near the equator and at middle latitudes.

The primary goal of our study is to pinpoint occasions when THEMIS and ERG traverse the same MLT sector within a relatively close time frame, enabling us to compare the characteristics of whistler-mode waves measured by these two missions. Instead of focusing on precise conjunction events \cite<e.g.,>{Colpitts20}, we rely on earlier observations of long-lasting domains of whistler-mode generation regions \cite{MartinezCalderon16,MartinezCalderon20}, which covers extensive spatial scales of approximately $\sim 1R_E$ and even larger scales \cite{Agapitov18:correlations}; here, $R_E$ denotes the Earth's radius. While the cross-field (equatorial) scales of individual generation regions usually do not exceed $\sim R_E/2$ \cite{Agapitov17:grl}, these source regions may occupy substantial MLT/$L$-shell sectors \cite{Agapitov18:correlations}. Long-living large-scale wave source regions are best seen in ground-based observations, which continuously monitor the wave activity within the same MLT/$L$-shell sectors  \cite{Titova15,Titova17,Demekhov20,Artemyev21:jgr:ducts,MartinezCalderon20}. Therefore, a comparison of multi-spacecraft whistler-mode wave observations can be performed with spacecraft measurements essentially separated in time.

To compare the intensities of THEMIS and ERG measured waves, we first need to confirm that these waves exhibit the same spectral characteristics. The most reliable and stable characteristic in this regard is the mean wave frequency, which varies by less than 50 percent between equatorial wave sources and latitudes exceeding $40^\circ$ \cite{Agapitov18:jgr}. Therefore, this characteristic is utilized in our study as a measure of wave spectrum similarity for THEMIS and ERG measurements. Two remaining characteristics of whistler-mode wave spectra, namely peak intensity and spectrum width, are expected to vary along the magnetic field line due to wave damping by Landau resonance with suprathermal electrons \cite{Bortnik07:landau,Chen13,Watt13:ray_tracing}, and wave ray divergence in inhomogeneous plasma \cite{Shklyar04,Breuillard13:angeo,Katoh14,Kang2024}. Therefore, the comparison of THEMIS and ERG measurements should reveal variability in these characteristics.

Empirical models predict that wave intensity increase away from the equator \cite<due to wave amplification within the source region, see>{Demekhov&Trakhtengerts08,Demekhov11,Katoh08:angeo,Katoh&Omura13,Omura21:review,Tao20,Tao21}, reach a maximum at middle latitudes, and then decay towards high latitudes \cite<see>{Agapitov15:jgr,Agapitov18:jgr}. However, such an empirical latitudinal profile has been constructed based on statistical wave measurements at different latitudes, and thus cannot exclude the presence of a wave population strongly localized around the equatorial source \cite<e.g., due to a large population of suprathermal electrons that quickly damp the waves away from the equator; see simulations in>{Chen13,Chen21:frontiers} or a wave population propagating to high latitudes without damping and wave ray divergence, due to wave ducting by plasma density fluctuations \cite{Hanzelka&Santolik19,Ke21:ducts,Chen21:ducting,Shen21:grl:ducts,Chen22:microbursts}. These two populations would be averaged out in statistical models but can be observed in events by nearby equatorial and mid-latitude spacecraft. Therefore, in this study, we will examine various wave populations with different latitudinal profiles of wave intensity.

The rest of this paper is organized as follows: in Section \ref{sec:data}, we describe the THEMIS and ERG datasets, spacecraft instrumentation, and data analysis technique. We also provide several example comparisons of the THEMIS/ERG wave spectra. In Section \ref{sec:stat}, we describe the main statistical characteristics of the THEMIS/ERG comparison. We discuss latitudinal variations of whistler-mode wave intensity and wave spectrum width, distinguishing between the two subsets of observations with different latitudinal profiles. Finally, in Section \ref{sec:discussion} we summarize our results and discuss their significance to modeling wave-particle resonant interactions within the radiation belts.

\section{Dataset and instruments}\label{sec:data}

In this study, we utilized near-equatorial measurements from THEMIS A spacecraft. The magnetic field, with a 3s (spin-rate) resolution, is provided by the flux-gate magnetometer \cite{Auster08:THEMIS}. The wave magnetic field spectra, covering 32 frequency channels (up to 8 kHz) at a 1s resolution \cite{Cully08:ssr}, are obtained from the search-coil magnetometer \cite{LeContel08}. We use the ${\it fff}$ dataset of onboard calculated wave spectra, but rescale it by $3/2$ to account for the fact that this dataset use only two from three magnetic field components. To facilitate a comparative analysis, we contrasted THEMIS measurements with those from the ERG spacecraft. The magnetic field, with an 8s resolution, is provided by the Magnetic Field Experiment onboard ERG \cite{Matsuoka18:ERG_MGF}. Additionally, the wave magnetic field spectra are provided by the wave instrument \cite{Kasahara18:ERG_PWE, Matsuda18}, covering the frequency range up to 20 kHz. To ensure compatibility, both THEMIS and ERG orbits are projected onto the MLT and $L$-shell domain.
$L$-shell values are calculated from Arase orbit Level-3 data files with \citeA{Tsyganenko&Sitnov05} model \cite{Miyoshi18:data}

\begin{figure}
    \centering
    \includegraphics[width=1\linewidth]{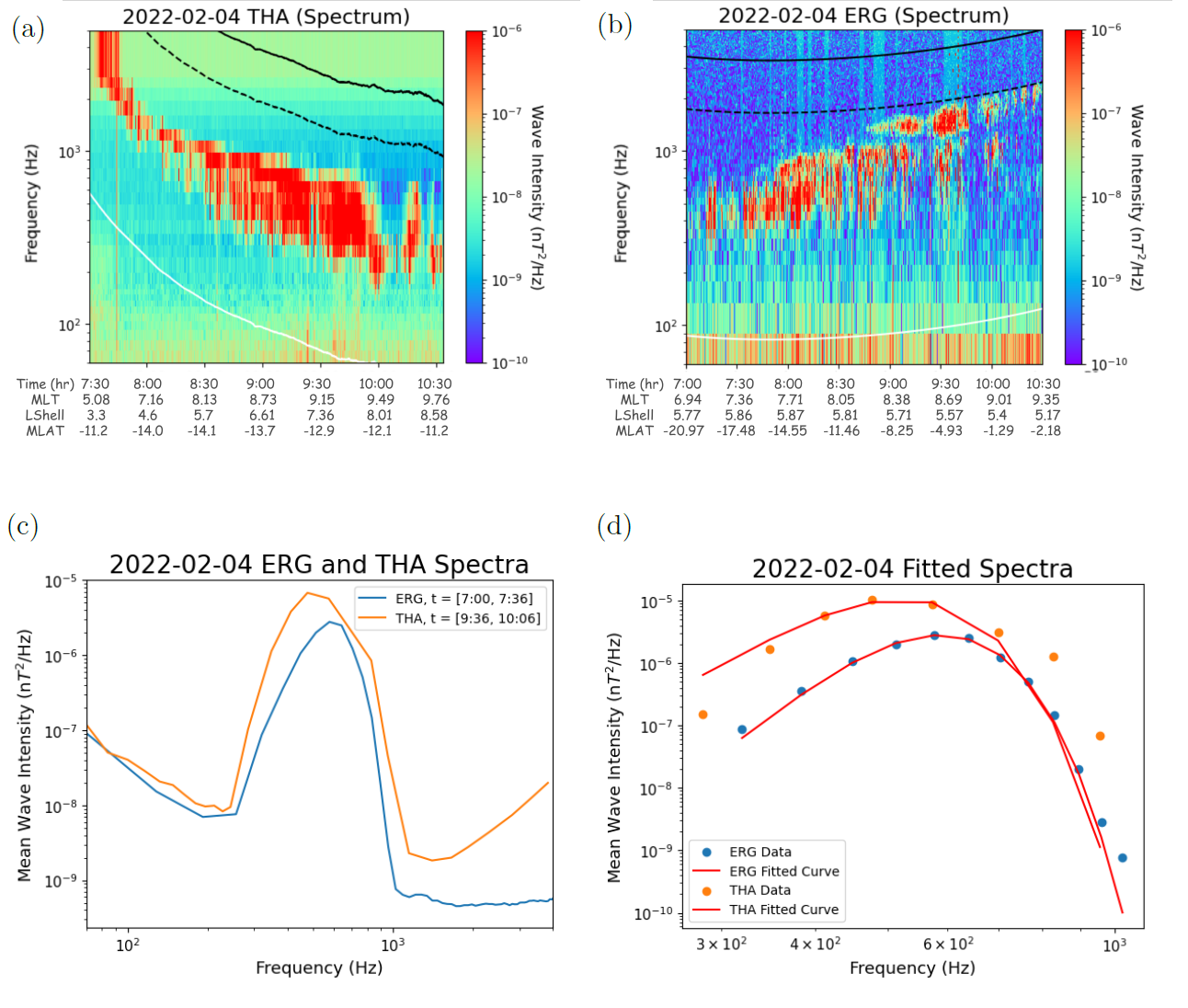}
        \caption{Overview of an example event on 04 February 2022: (a) dynamical spectrum of the THEMIS magnetic field, (b) dynamical spectrum of the ERG magnetic field, (c) comparison of THEMIS and ERG spectra averaged over intervals indicated in Table \ref{table}, (d) THEMIS and ERG spectrum peaks and their fittings by the Gaussian function.
        Black lines in (a,b) show electron gyrofrequency and its fractions (for ERG, we use the magnetic field projected to the equatorial plane).
        Note that we rescaled the THEMIS {\it fff} data by a factor of 3/2 in order to compare with ERG data (see text for details).}
        \label{fig:event01}
\end{figure}

We primarily focus on the time interval between January and April 2022, which is characterized by numerous THEMIS/ERG crossings within the same MLT and $L$-shell sectors. Additionally, we incorporated several events from November 2019 and May-June 2021. The comprehensive list of events is shown in Table \ref{table}, where we also provide the average MLT values for both spacecraft and the average $MLAT$ for ERG. Figure \ref{fig:event01} illustrates sample events from this table. Panels (a, b) depict magnetic field spectra for THEMIS and ERG. The whistler-mode waves are clearly visible between the lower hybrid (white curve) and half of the electron cyclotron (dashed black curve) frequencies. The wave frequency range aligns with the magnetic field variation along the spacecraft orbits. We selected sub-intervals for THEMIS from 09:30UT to 10:00UT and for ERG from 07:30UT to 08:00UT when both spacecraft were within the same MLT range of $[7.5,10]$ and $L$-shell range of [5.7,8]. Panel (c) compares the average wave spectra for THEMIS (orange) and ERG (blue) during the specified sub-intervals. A distinct peak is observed between 200 Hz and 1 kHz by both spacecraft. To determine the main wave spectral characteristics (peak intensity, spectrum width, and mean frequency), we fit the wave spectra to a Gaussian function. This fitting approach is conventionally used to assess the impact of whistler-mode waves on radiation belt electrons \cite{Glauert&Horne05}.
\[
B_w^2 \left( f \right) = A\exp \left( { - \frac{{\left( {f - f_0 } \right)^2 }}{{\delta f^2 }}} \right)
\]

Panel (d) displays the fitting for THEMIS and ERG spectra, with the corresponding parameter values as follows: $A_{THM}=6.7\cdot10^{-6} {\rm nT^2/Hz}$ and $A_{ERG}=2.8\cdot10^{-6} {\rm nT^2/Hz}$, $f_{0,THM}=537{\rm Hz}$ and $f_{0,ERG}=586{\rm Hz}$, $\delta f_{THM}=150 {\rm /Hz}$ and $\delta f_{ERG}=137{\rm Hz}$. The mean wave frequencies are nearly identical for both spectra, and the spectrum widths are also quite close. This consistency confirms that we are comparing waves generated within the same plasma by hot electrons with similar distribution functions. However, the peak wave intensities differ, with waves observed at ERG (located at $\langle MLAT \rangle \approx 18^\circ$) being weaker than the near-equatorial waves observed by THEMIS.

\begin{figure}
    \centering
\includegraphics[width=1\linewidth]{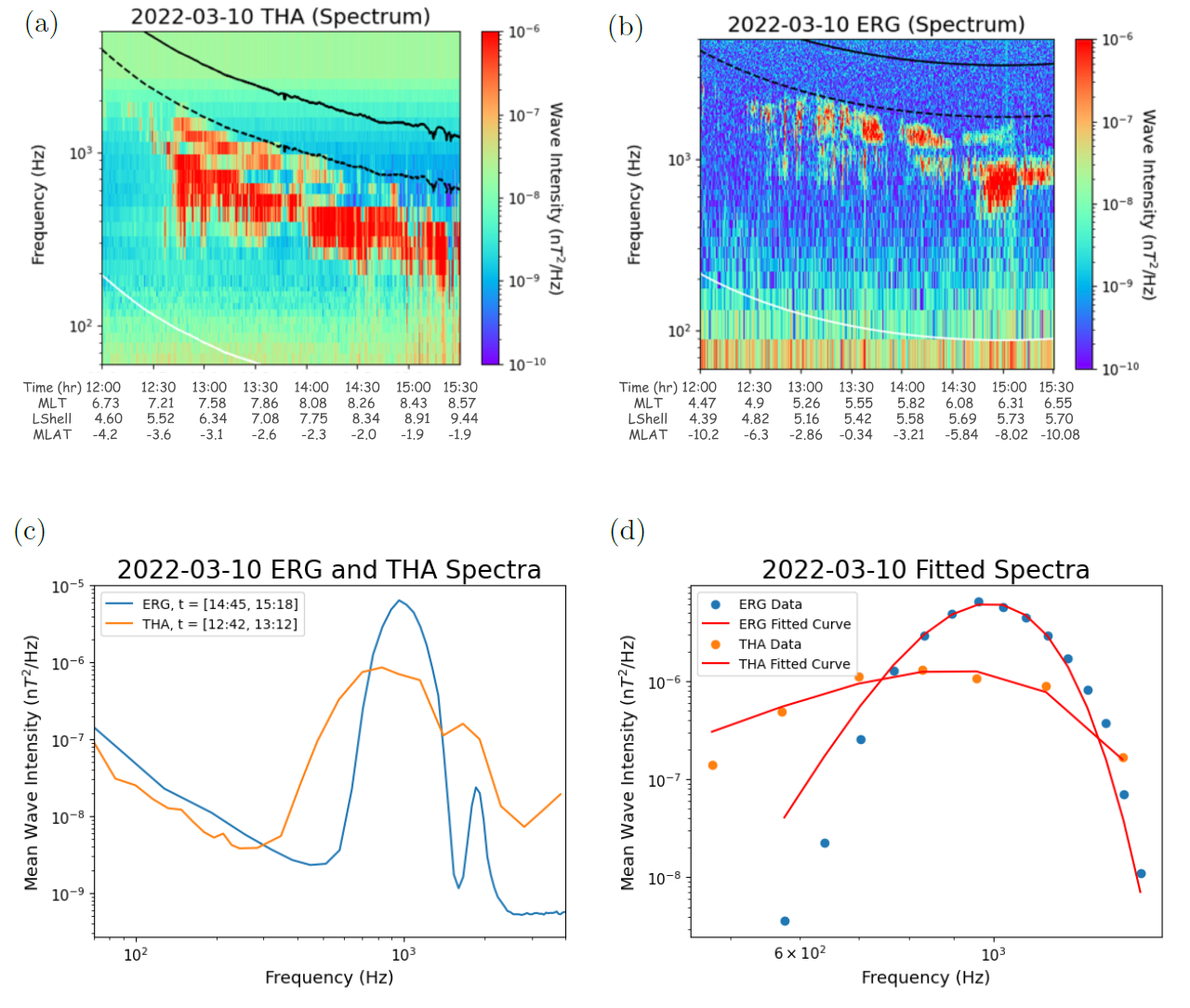}
        \caption{Overview of an example event on 10 March 2022: (a) dynamical spectrum of the THEMIS magnetic field, (b) dynamical spectrum of the ERG magnetic field, (c) comparison of THEMIS and ERG spectra averaged over intervals indicated in Table \ref{table}, (d) THEMIS and ERG spectrum peaks and their fittings by the Gaussian function.
        Black lines in (a,b) show electron gyrofrequency and its fractions (for ERG, we use the magnetic field projected to the equatorial plane).
        Note that we rescaled the THEMIS {\it fff} data by a factor of 3/2 to compare with ERG data (see text for details).}
        \label{fig:event02}
\end{figure}

Figure \ref{fig:event02} illustrates an example event wherein THEMIS and ERG spectra exhibit a more statistically typical increase in wave intensity away from the equator. Panels (a) and (b) depict the wave spectra measured by THEMIS and ERG, respectively. We specifically focus on the interval from 12:45 to 13:15 UT for THEMIS and from 14:45 to 15:15 UT for ERG, during which both spacecraft traversed the same MLT and $L$-shell domain. Panel (c) confirms the similarity of the measured spectra by THEMIS and ERG for these intervals.

Despite a two-hour delay between THEMIS and ERG measurements, the wave spectra exhibit notable similarity, including a double peak as shown in panel (c) and the main peak frequency, which is better observed in panel (d). This consistency affirms the stability of the wave source region. The wave intensity measured by THEMIS is approximately four times smaller than the intensity measured by ERG at $MLAT\sim 10^\circ$ (see panel (d)). Thus, this event substantiates general statistical findings regarding wave amplification away from the equatorial source region \cite{Agapitov18:jgr}.

The complete list of events for our statistical analysis is presented in Table \ref{table}. We have 41 intervals of THEMIS and ERG wave measurements within the same $MLT$ and $L$-shell sector. These events exhibit a wide range of latitudinal and temporal spacecraft separations, allowing us to examine the dependence of wave intensity on latitudes and explore the timescales over which such dependence holds. The majority of intervals are from the dawn-to-noon sector, where whistler-mode waves are typically observed well off the equator \cite{Meredith12,Agapitov13:jgr}. Additionally, there is a smaller subset of intervals from the pre-midnight sector, where whistler-mode waves are generally confined around the equatorial plane \cite{Meredith12,Agapitov13:jgr}.

 \begin{table}
 \caption{List of events with THEMIS and ERG observations of whistler-mode waves.}
\begin{tabular}{| l | c | c | c | c | c |}
 date & ERG           & THEMIS        & ERG                   & THEMIS                & ERG\\
      & time interval & time interval & $\langle MLT \rangle$ & $\langle MLT \rangle$ & $\langle MLAT \rangle$\\
 \hline
2019-11-30 & 07:12-07:48& 07:18-07:48& 4.1  &5.6  &24.5 \\
2019-12-12 & 22:33-22:45& 21:48-22:24& 4.3  &7.7  &38.7 \\
2021-04-30 & 03:06-03:42& 04:50-05:06& 1.6  & .4  &13.7 \\
2021-05-26 & 10:54-11:30& 09:48-10:24& 22.4 &22.9 &24.8 \\
2021-06-02 & 04:30-05:24& 05:47-05:51& 22.7 &23.3 & 1.2 \\
2021-06-05 & 17:18-18:00& 16:00-16:24& 22.3 &22.2 &20.5 \\
2021-06-06 & 20:42-21:06& 19:42-20:06& 21.6 &23.8 &9.2 \\
2021-06-13 & 14:00-15:00& 14:30-15:00& 21.7 &21.4 &39.5 \\
2022-01-02 & 07:18-07:27& 09:21-09:42& 10.3 &12.1 &22.3 \\
2022-01-03 & 10:00-11:24& 11:18-12:24& 9.7  &11.4 &18.2 \\
2022-01-11 & 10:30-11:24& 10:18-10:42& 11.6 &10.3 &3.2 \\
2022-01-13 & 18:06-19:00& 16:36-17:36& 10.3 &10.1 &3.7 \\
2022-01-18 & 08:00-09:00& 06:24-07:00& 7.8  &9.7  &22.7 \\
2022-01-21 & 16:36-17:00& 15:54-16:36& 10.8 &9.5  &11.5 \\
2022-01-22 & 19:12-19:42& 19:42-20:06& 10.3 & 8.8 &7.7 \\
2022-01-27 & 12:06-12:36& 11:18-11:42& 9.5  & 9.8 &2.8 \\
2022-01-29 & 14:06-15:00& 12:06-13:00& 11.7 &16.3 &20.1 \\
2022-01-30 & 19:18-19:27& 18:06-18:30& 12.2 & 8.1 &30.6 \\
2022-02-02 & 02:54-03:12& 00:24-00:54& 10.2 &6.4  &2.3 \\
2022-02-03 & 05:00-05:30& 04:54-05:12& 8.6  &8.4  &14.3 \\
2022-02-04 & 07:00-07:36& 09:30-10:06& 7.2  &9.8  &18.4 \\
2022-02-05 & 11:27-11:48& 12:00-12:24& 7.3  & 9.2 &9.6 \\
2022-02-08 & 20:24-20:48& 20:30-20:54& 11.3 &6.7  &25.2 \\
2022-02-11 & 02:48-03:12& 03:54-04:06& 8.5  &7.6  &10.3 \\
2022-02-12 & 06:06-06:18& 08:00-08:12& 7.9  &8.5  &10.7 \\
2022-02-13 & 10:15-10:27& 11:30-11:48& 7.9  &8.9  & 0.7 \\
2022-02-20 & 05:30-06:00& 05:48-06:12& 9.5  & 5.8 &7.4\\
2022-02-21 & 09:48-10:30& 10:00-10:18& 9.7  &7.6  &16.4 \\
2022-02-23 & 17:24-17:42& 16:48-17:12& 8.0  &7.9  &4.3 \\
2022-03-04 & 18:12-19:00& 19:03-19:33& 7.4  &7.0  &5.6 \\
2022-03-07 & 02:06-02:30& 02:24-02:54& 6.6  & 7.2 &7.6 \\
2022-03-10 & 14:45-15:18& 12:42-13:12& 6.3  & 7.5 &8.1\\
2022-03-11 & 16:24-16.33& 15:30-15:39& 4.6  & 7.0 &21.1 \\
2022-03-17 & 10:42-11:06& 08:36-09:24& 9.0  & 7.0 &29.4\\
2022-03-25 & 07:06-07:30& 06:45-07:00& 7.8  & 5.2 &22.1\\
2022-03-26 & 09:36-09:48& 11:06-11:36& 6.0  & 6.5 &20.0 \\
2022-04-11 & 09:21-09:36& 09:12-09:24& 2.9  & 5.0 &5.2\\
2022-04-18 & 04:54-05:12& 05:18-05:30& 4.6  & 4.7 &21.5 \\
2022-04-19 & 08:18-08:42& 08:48-09:12& 3.8  & 5.0 &26.5\\
2022-04-21 & 14:00-14:18& 14:00-14:30& 3.3  & 4.5 &18.9 \\
\hline
\end{tabular}
 \label{table}
\end{table}

\section{Statistical results}\label{sec:stat}

Two events depicted in Figs. \ref{fig:event01} and \ref{fig:event02} are representative of our statistical findings listed in Table \ref{table}: both wave amplification and damping in the off-equatorial region are observed. However, due to a substantial time delay between spacecraft measurements for some events (up to about $2$ hours), comparing the wave intensities measured by THEMIS and ERG may only be meaningful if we demonstrate that other characteristics of the wave spectra (associated with the properties of the wave source region) remain relatively unchanged. Therefore, our first step is to compare these characteristics.

\begin{figure}
\includegraphics[width=1\linewidth]{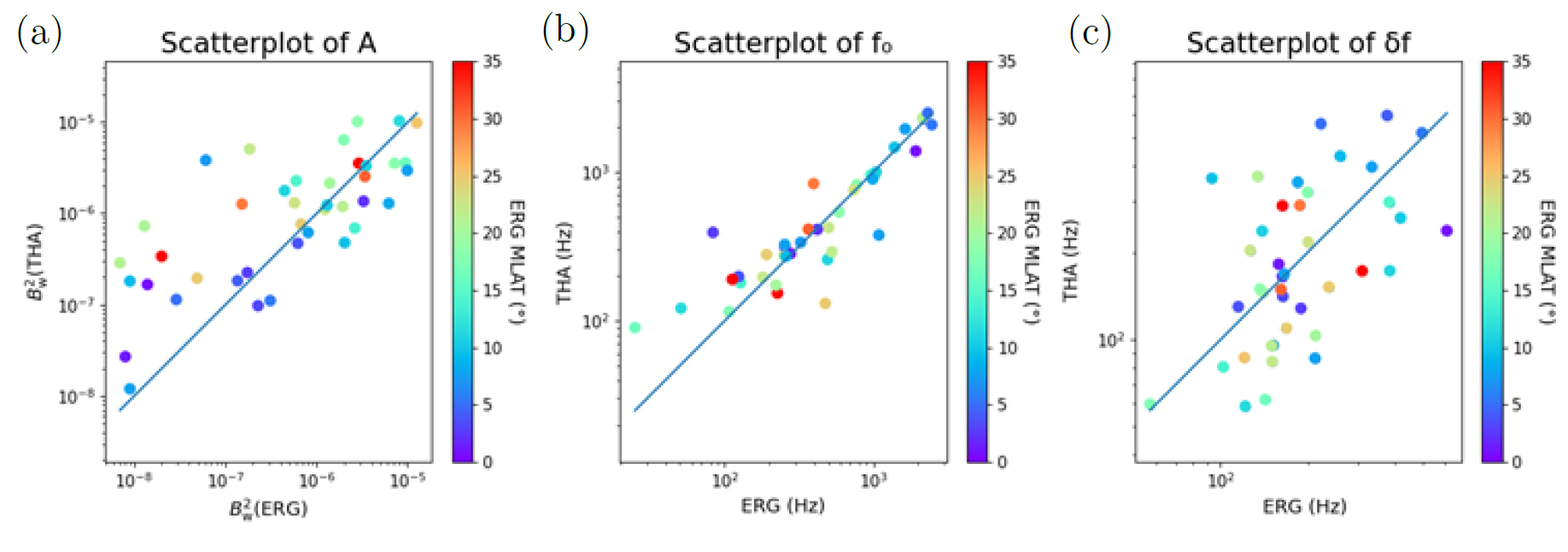}
    \caption{Statistical comparison of whistler-mode spectrum characteristics during ERG and THEMIS conjunction events: (a) peak wave intensity, (b) mean wave frequencies, and (c) widths of wave spectra. Color codes the magnetic latitude of ERG observations. }
        \label{fig:statistics_mlat}
\end{figure}

Figure \ref{fig:statistics_mlat} illustrates a comparison of wave intensities, mean wave frequencies, and widths of the wave spectra. The color in the plot represents the ERG magnetic latitude, indicating the distance of ERG from THEMIS along the magnetic field line. While the frequency width may vary significantly between ERG and THEMIS, the mean frequencies mostly follow a 1:1 trend. Moreover, the correlation of wave intensities also demonstrates interesting results. There are two distinguished groups of events: small wave intensity at ERG generally means larger wave intensity at THEMIS, whereas large wave intensity at ERG is more often associated with smaller wave intensity on THEMIS. The first group of events show wave damping away from the equatorial region, and the second group of events show possible wave amplification.

As shown in Fig. \ref{fig:statistics_mlat}, there is no clear dependence of the relation between THEMIS and ERG parameters on the ERG magnetic latitude: large deviations in wave frequencies and intensities between THEMIS and ERG observations can occur for small or large magnetic latitudes. Therefore, a more detailed analysis is needed to identify the subset of intervals that exhibit an increase in wave intensity away from the equator.

\begin{figure}
\includegraphics[width=1\linewidth]{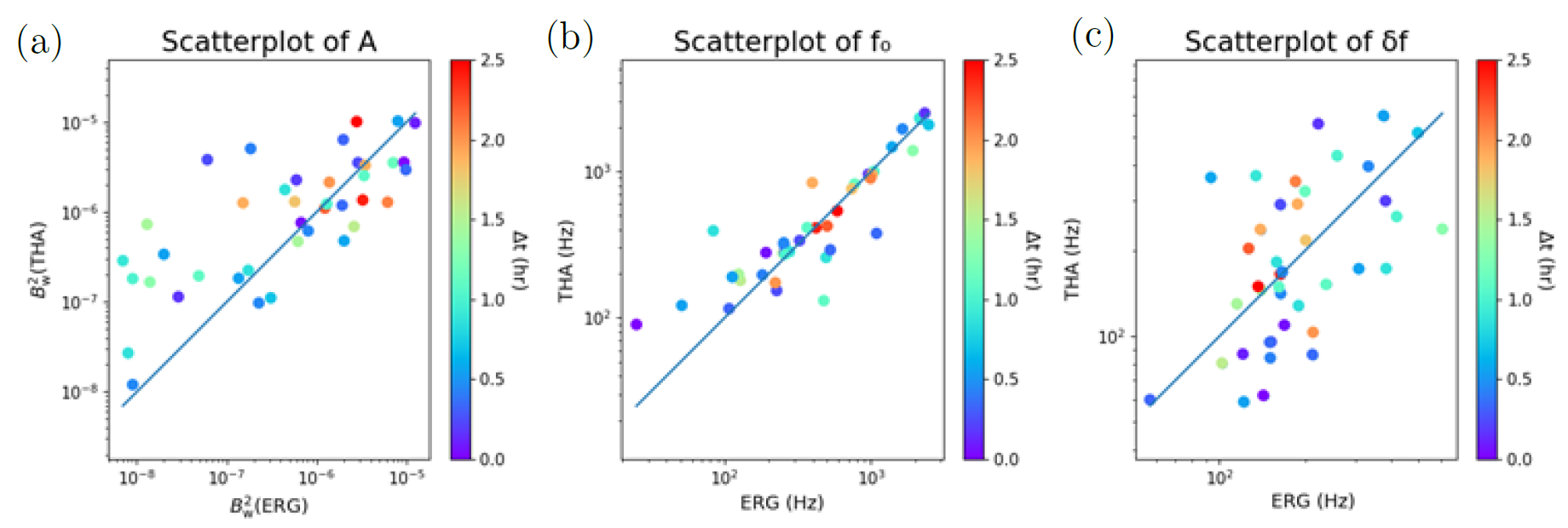}
\caption{Statistical comparison of whistler-mode spectrum characteristics during ERG and THEMIS conjunction events: (a) peak wave intensity, (b) mean wave frequencies, and (c) widths of wave spectra. The color in the figures codes the time delay between THEMIS and ERG observations.}
        \label{fig:statistics_dt}
\end{figure}

Figure \ref{fig:statistics_dt} is the same as Fig. \ref{fig:statistics_mlat}, but with color coded by the time separation between THEMIS and ERG observations. Correlation of THEMIS and ERG wave amplitudes, as well as mean frequencies, does not show any dependence on the time delay between THEMIS and ERG. This further confirms that all selected intervals contain quasi-stationary wave source regions with unchanged properties within the time-scale including THEMIS and ERG observations, and difference in observed wave intensity by THEMIS and ERG is due to different magnetic latitudes, rather than temporal variability.

\begin{figure}
        \centering
\includegraphics[width=1\linewidth]{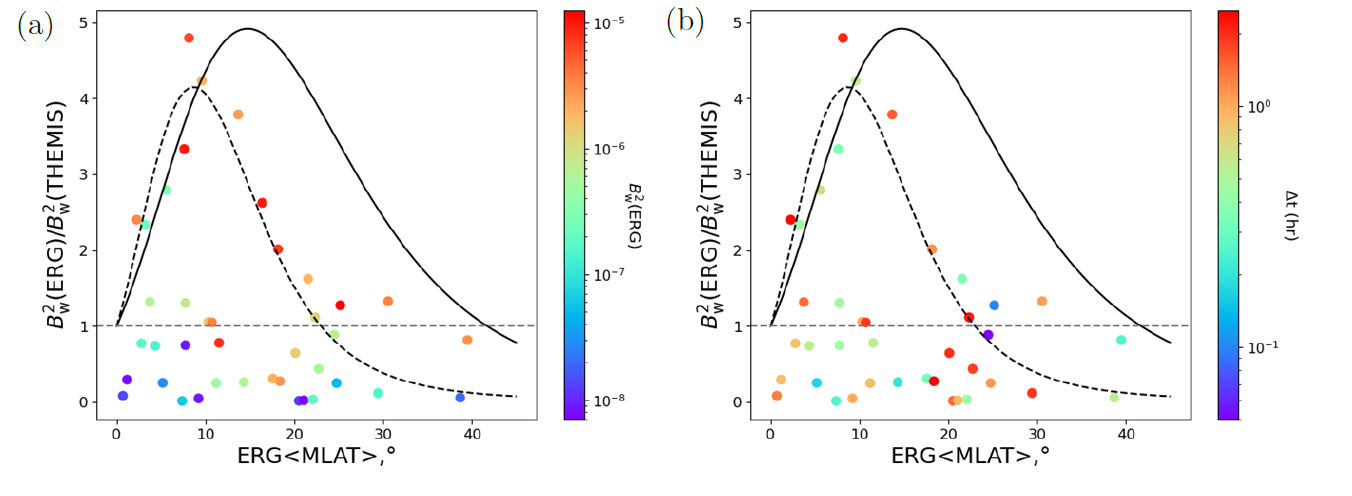}
 \caption{Comparison of the ratio of ERG and THEMIS wave intensity with the prediction of \cite{Agapitov18:jgr} model. Dashed and solid black lines show model profiles for moderate and high geomagneti activity levels.  Data points are color-coded by $\Delta MLT$(a) and $\Delta t$(b).\label{fig:comparison}    }
\end{figure}

To compare THEMIS and ERG observations with the empirical wave model \cite{Agapitov18:jgr}, we plotted the ratio of average wave intensities $A(ERG)/A(THEMIS)$ as a function of the ERG magnetic latitude. This representation demonstrates the variation of wave intensity normalized to the equatorial value along the magnetic field line. Figure \ref{fig:comparison}(a) illustrates how this data representation aids in distinguishing between two types of events.

For events where intense waves are observed by ERG, the ratio $A(ERG)/A(THEMIS)$ conforms to the expected model profile: it increases with magnetic latitudes, reaches a maximum value around 10-20 degrees, and then decreases. Conversely, for events with low-intensity waves observed by ERG, the ratio $A(ERG)/A(THEMIS)$ is typically below one, suggesting that these waves likely experience strong Landau damping as they propagate away from the equator \cite{Bortnik08,Chen13,Watt13:ray_tracing}. It is worth noting that despite their low intensity, the waves exhibit similar mean frequencies in THEMIS and ERG, indicating that they are indeed generated at the equator and propagate to ERG latitudes with conserved frequency but reduced amplitudes. Importantly, Fig. \ref{fig:comparison}(b) confirms that the ratio $A(ERG)/A(THEMIS)$ does not depend on the time delay between THEMIS and ERG measurements. Both event types, with $A(ERG)/A(THEMIS)$ following the empirical model and with $A(ERG)/A(THEMIS)<1$ for all latitudes, can be observed even with a delay of less than one hour.

\section{Discussion and conclusions}\label{sec:discussion} We conducted a multi-event comparison of whistler-mode wave spectra measured by the near-equatorial THEMIS and off-equatorial ERG spacecraft. Each selected event includes THEMIS and ERG measurements within the same MLT sector (see table \ref{table}) with a time delay of less than $2$ hours. While these events offer a less rigorous equatorial/off-equatorial matching compared to meticulously-selected conjunction events \cite<as shown in>{Colpitts20}, they constitute a significant statistics and provide higher accuracy than global wave statistics that mixes wave measurements spanning different months and years to encompass a broader latitudinal range \cite<e.g.,>{Meredith12, Agapitov13:jgr}.

The primary conserved characteristic of whistler-mode waves is the mean frequency of the wave spectrum, which changes along magnetic field lines primarily due to preferential damping of the higher frequency part of the spectrum \cite<see discussion in>{Agapitov18:jgr}. We predominantly examine data on the dawn/day sector where weaker Landau damping effects are expected \cite<see discussion in>{Kang2024}, and utilize the mean wave frequency to demonstrate that THEMIS and ERG capture the same wave spectra. Indeed, Figs. \ref{fig:statistics_mlat} and \ref{fig:statistics_dt}(b) illustrate a strong correlation between THEMIS and ERG mean wave frequencies, validating the event selection process.

Despite the relatively good correlation of wave frequencies between measurements from THEMIS and ERG, we observe a large population of waves with off-equatorial wave intensity (ERG) smaller than the near-equatorial wave intensity (THEMIS). These events are characterized by weak waves at ERG (see Fig. \ref{fig:comparison}(a)).
We have further confirmed that these events are characterized by close values of THEMIS and ERG mean wave frequencies, suggesting that THEMIS and ERG likely observed waves from the same source region but with different intensities. Although Fig. \ref{fig:comparison}(b) does not reveal any correlation between the time delay of THEMIS/ERG observations and wave intensities, it is worth noting that most events where ERG observed low-intensity waves at small latitudes ($<10^\circ$) are associated with $\Delta t \geq 0.5$ hours. This time span is sufficiently long to potentially alter the properties of the wave generation region \cite<see discussion of time-scales of the wave generation variability in>{Tao11}. Excluding this population of low-intensity waves observed at ERG, the dataset of THEMIS/ERG observations follows the empirical model profile.

Verification of empirical wave models is important because these models of wave intensity distribution across spatial (MLT, $L$-shell, MLAT) domains rely on multi-spacecraft measurements at different times, under the assumption of the same geomagnetic activity levels. This approach assumes a single dominant wave population originating at the equator and propagating poleward, shaping the latitudinal profile of wave intensity. However, this profile could potentially result from a combination of two populations: (1) almost undamped, ducted waves maintaining consistent intensity across broad latitudes, and (2) strongly damped waves localized near the equator. Different proportions between these two populations could explain any observed latitudinal profile of the wave intensity. Therefore, an independent verification of the single-population hypothesis is crucial. Our combined THEMIS and ERG measurements demonstrate that waves observed at different latitudes within one hour indeed align with the latitudinal wave intensity profile from the empirical model, underscoring the need for further verification in the future.

\acknowledgments
A.V.A. and X.J.Z. acknowledge support by NASA awards 80NSSC20K1270, 80NSSC23K0403, 80NSSC24K0558, NAS5-02099, and NSF grant AGS-2021749.

\section*{Open Research}

\noindent THEMIS data is available at http://themis.ssl.berkeley.edu.

\noindent Science data of the ERG (Arase) satellite were obtained from the ERG Science Center operated by ISAS/JAXA and ISEE/Nagoya University (https://ergsc.isee.nagoya-u.ac.jp/index.shtml.en, \cite{Miyoshi18:center}). The present study analyzed the MGF L2\_v04\_04 data \cite{Matsuoka18:data}, PWE OFA L2\_v02\_01 data \cite{Kasahara18:data}, and ORB L2\_v03 data \cite{Miyoshi18:data}.
ERG $L$-shell values are obtained from \url{https://www.isee.nagoya-u.ac.jp/doi/10.34515/DATA_ERG-12001.html}

\noindent Data access and processing were performed using SPEDAS V3.1, see \cite{Angelopoulos19}.


\end{document}